\documentclass[runningheads]{llncs}

\renewcommand{\qed}{\hfill\rule{1ex}{1ex}}

\usepackage{amssymb}
\usepackage{amsmath}
\usepackage{bbm}
\usepackage{float}
\usepackage{graphicx}
\usepackage{amsfonts}
\usepackage{amssymb}
\usepackage{paralist}
\usepackage{floatflt}
\usepackage{alltt}
\usepackage{color}
\usepackage{algorithm}
\usepackage{algorithmic}
\usepackage{hyperref}
\usepackage{xcolor}
\definecolor{string}{rgb}{0.7,0.0,0.0}
\definecolor{comment}{rgb}{0.13,0.54,0.13}
\definecolor{keyword}{rgb}{0.0,0.0,1.0}

\usepackage{tikz}

\newcommand{\x}{\ensuremath{x}}
\newcommand{\y}{\ensuremath{y}}
\newcommand{\s}{\ensuremath{s}}
\newcommand{\w}{\ensuremath{w}}
\newcommand{\z}{\ensuremath{z}}

\newcommand{\snought}{\ensuremath{s_0}}

\newcommand{\Us}{\ensuremath{U^s}}
\newcommand{\Uy}{\ensuremath{U^y}}
\newcommand{\Ux}{\ensuremath{U^x}}
\newcommand{\Uyvt}{\ensuremath{U^y_{v,t}}}
\newcommand{\Uxvt}{\ensuremath{U^x_{v,t}}}
\newcommand{\Lst}{\ensuremath{L^s_t}}
\newcommand{\Lyvt}{\ensuremath{L^y_{v,t}}}
\newcommand{\Lxvt}{\ensuremath{L^x_{v,t}}}

\newcommand{\xt}{\ensuremath{x_t}}
\newcommand{\yt}{\ensuremath{y_t}}
\newcommand{\xvt}{\ensuremath{x_{v,t}}}
\newcommand{\yvt}{\ensuremath{y_{v,t}}}
\newcommand{\sst}{\ensuremath{s_t}}
\newcommand{\wt}{\ensuremath{w_t}}
\newcommand{\zt}{\ensuremath{z_t}}
\newcommand{\wvt}{\ensuremath{w_{v,t}}}
\newcommand{\zvt}{\ensuremath{z_{v,t}}}
\newcommand{\pt}{\ensuremath{p_t}}

\newcommand{\Rt}{\ensuremath{R_t}}

\newcommand{\Ust}{\ensuremath{U^s_t}}

\newcommand{\tbound}{\ensuremath{T}}
\newcommand{\tset}{\mathcal{T}}
\newcommand{\vset}{\mathcal{V}}
\newcommand{\vbound}{\ensuremath{V}}

\newcommand{\compset}[2]{\mathcal{C}_{#1,#2}}
\newcommand{\cvarone}{\ensuremath{V_1}}
\newcommand{\cvartwo}{\ensuremath{V_2}}
\newcommand{\cboundone}{\ensuremath{B_1}}
\newcommand{\cboundtwo}{\ensuremath{B_2}}
\newcommand{\skipthis}[1]{}
\newtheorem{assume}[definition]{Assumption}

\begin{document}

\title{Warehouse Problem with Multiple Vendors and Generalized Complementarity Constraints
\thanks{This work  was partially supported by ONR Grant N00014-21-1-2575.}
}
\titlerunning{Warehouse Problem}
\author{Ishan Bansal\inst{1}\and Oktay G\"unl\"uk\inst{2}}
\authorrunning{I. Bansal, O. G\"unl\"uk}
\institute{Cornell University,   \email{ib332@cornell.edu} \\\and Cornell University,   \email{ong5@cornell.edu}}
\maketitle
\begin{abstract}
    We study the warehouse problem, arising in the area of inventory management and production planning. Here, a merchant wants to decide an optimal trading policy that computes quantities of a single commodity to purchase, store and sell during each time period of a finite discrete time horizon. Motivated by recent applications in energy markets, we extend the models by Wolsey and Yaman (2018) and Bansal and Günlük (2023) and consider markets with multiple vendors and a more general form of the complementarity constraints. We show that these extensions can capture various practical conditions such as surge pricing and discounted sales, ramp-up and ramp-down constraints and batch pricing. We analyze the extreme points of the underlying non-linear integer program and provide an algorithm that exactly solves the problem. Our algorithm runs in polynomial time under reasonable practical conditions. We also show that the absence of such conditions renders the problem NP-Hard.
    \keywords{Warehouse Problem \and Inventory Management \and Algorithm Design \and Polyhedral Analysis \and Extreme Points \and Electricity Trading}
\end{abstract}

\section{Introduction}

Inventory management and production planning occupy a central role in the study and analysis of market dynamics. In this paper we study the warehouse problem which considers the decisions that a merchant makes while trading a particular product and maintaining a storage warehouse. The merchant seeks to leverage price fluctuations to maximize their profits and aims to compute optimal quantities of the product to purchase, sell and store during each time period of a finite discrete time horizon. The problem was first introduced by Cahn \cite{Cahn} as follows: {\em``Given a warehouse with fixed capacity and an initial stock of a certain product, which is subject to known seasonal price and cost variations, what is the optimal pattern of purchasing (or production), storage and sales?"}. The problem is closely related to and is a generalization of the more well-known lot-sizing problem, differing primarily in the decisions related to sales. In the classical lot-sizing problem, the merchant has to adhere to a fixed demand and cannot choose to sell more or less than this fixed demand. In contrast, the warehouse problem takes a more flexible approach where sales can be penalized or constrained as desired.

Since its introduction, the warehouse problem has found applications in the commercial management of inventories in commodity markets \cite{secomandi,devalkar,wu2012,qin2012,faghih2013,zhou2016,zhou2019,halman2018,bolun2021,xuconference}. These include the management of conventional warehouses to store commodities such as agricultural products as well as management of tanks and reservoirs to store oil, water and gas. More recently, the problem and its variants have received attention due to their direct applications in energy markets. In these markets, both merchants and consumers have the ability to buy and sometimes generate electricity, storing it using various devices, and subsequently redistributing it to serve other clients or selling it back to the grid. Electricity prices are known to be notoriously volatile \cite{anderson2008,conejo2010} creating opportunities for arbitrage \cite{graves1999opportunities,peterson2010economics} that people can benefit from, if provided with the right set of tools. The emerging relevance of renewable energies in electricity markets \cite{su2013modeling,gast2014optimal,zhou2019} has also been a contributing factor to the surge of interest in the warehouse problem.

The warehouse problem addresses practical decisions pertaining to inventory management, hence a primary focus of research has been to consider additional constraints and properties that model real-world applications more accurately. The original warehouse problem introduced by Cahn \cite{Cahn} considered deterministic and linear cost and sales prices and unbounded purchase and sales capacities. This was solved using linear programming by \cite{charnes1955}, using dynamic programming by \cite{bellman} and using analytical methods by \cite{dreyfus}. A model with stochastic and uncertain costs was introduced by \cite{charnes} and such models are often represented using a Markov decision process. A common practical approach to solve such stochastic problems is the use of deterministic models with re-optimization at successive time steps \cite{lai2010approximate,wu2012,secomandi,secomandi2015merchant}. This approach encourages further research on deterministic models. Other extensions include continuous-time models \cite{kaminski2008}, bounds on purchases \cite{rempala1994optimal}, time-independent lower and upper bounds on both purchases and sales \cite{secomandi}, multiple commodities and vendors \cite{devalkar}, non-linear objective functions \cite{xujournal}, negative pricing \cite{zhou2016}, fixed costs \cite{wolsey} and complementarity constraints that prohibit purchases and sales at the same time \cite{wolsey}. A Lagrangian approach to the problem was considered by \cite{cruise2019}. Recently, Bansal and Gunluk \cite{bansal2023} combined some of the above extensions into one model and provided conditions under which the problem goes from being NP-Hard to polynomial-time solvable. They provided first known polytime algorithms in settings with time-dependent bounds on purchases, and sales, fixed costs and complementarity constraints. This is the starting point of our work.

Motivated by applications to energy and electricity markets, we extend the model of \cite{bansal2023} by considering a market with multiple vendors and by generalizing the complementarity constraints that permit the decisions to produce/sell to interact and affect decisions in future time periods as well. While warehouse models with multiple vendors have been considered in the past \cite{devalkar}, these models generally make assumptions like linear costs that help reduce the multi-vendor instance to a single vendor one. We do not make such assumptions. To the best of our knowledge, the complementarity constraints have not been generalized in past literature. We show that these extensions permit us to model practical constraints like surged costs and discounted sales prices \cite{barreto2015incentives}, production ramp-up and ramp-down constraints \cite{wang2016value}, and batch pricing \cite{li2004dynamic}. We analyze this new model by providing conditions under which the problem becomes polynomial time solvable and provide the first known polynomial time algorithms in these settings. The conditions, in our opinion, are practical. For instance, it is reasonable to assume that the number of vendors in the market is `small' since merchants do not trade with too many vendors simultaneously. Additionally, we assume that the time interval between two production/sales decisions that affect each other is `small'. Our notion of small is logarithmic, and not necessarily constant. We supplement these results by showing that the absence of such conditions renders the problem $NP$-Hard. 

We begin by providing a characterization of the extreme points of the convex hull of the feasible region. To the best of our knowledge, such a characterization has not been previously provided for the warehouse problem with multiple vendors. To do this, we first analyze the inventory levels in extreme points (similar to \cite{wolsey,bansal2023}) as the production and sales quantities can then be analyzed in each time period separately. The goal is to use this characterization and construct a dynamic programming style network flow formulation that aids us in searching through the exponentially sized set of extreme points quickly. However, unlike \cite{wolsey,bansal2023}, the generalized complementarity constraints make this step a bit more nuanced, with the need for storing additional information in each state of the dynamic program. In particular, we store complementarity constraints that need to be satisfied in the future.

\section{Problem Formulation}

We begin by describing the model introduced in \cite{bansal2023} (which extends the model introduced in \cite{wolsey} by considering bounds on purchases and sales) and then explain the extensions we make to this model. Let $\tset = \{1,\ldots,\tbound\}$ denote the planning periods. Let $\Lst,L^x_t, L^y_t$ and $\Ust,U^x_t,U^y_t$ be nonnegative lower and upper bounds on storage, purchase and sales quantities respectively during time period $t$. Let variables $s_t,x_t$ and $y_t$ denote the stock level at the end of time period $t$, and the purchase and sales quantities during time period $t$ respectively. In addition, for any time period $t\in\tset$, define a binary variable $\wt$ which can take value zero only if no purchases were made during time period $t$, 
and a binary variable $\zt$ which can take value zero only if no sales were made. Then the model in \cite{bansal2023} can be expressed as the following mixed integer program.

\begin{align*}
    \max & \;\;\;\;\mathbb{E}_R\left[\sum_{t\in\tset} \pt(\xt,\yt,\sst,\wt,\zt \;|\; \Rt)\right]\hskip5cm\\
\end{align*} \begin{align*}
\text{s.t.}&\;\;\;\; \sst = \s_{t-1} - \yt + \xt & \forall\; t\in\tset\\
    &\;\;\;\;\xt\yt = 0 &\forall\; t\in\tset\\
    &\;\;\;\; y_t \leq s_{t-1} &\forall\; t\in\tset\\
    &\;\;\;\; L^s_t\leq \sst\leq \Us_t,\;\;\;\;L^y_tz_t\leq \yt\leq \Uy_t\zt,\;\;\;\;L^x_t w_t\leq \xt\leq \Ux_t\wt& \forall\; t\in\tset \\
    &\;\;\;\; \wt,\zt \in \{0,1\} &\forall\; t\in\tset
\end{align*}
where $\snought$ is an input parameter which denotes the initial stock at the beginning of the planning horizon, and $\pt(\xt,\yt,\sst,\wt,\zt \;|\; \Rt)$ denotes the pay-off at time period $t$ as a function of the decision variables.
Here $\Rt$ is a random variable that models the state of the market at time period $t$. This could include information on the selling and buying prices, holding costs and fixed costs for example. In most applications, the state of the market $R = \{\Rt\}_{t\in\tset}$ evolves according to a Markov process. At time period $t$, the state of the market $\Rt$ is realized and becomes known to the merchant. The binary variables $\wt$ and $\zt$ are used to model fixed costs, and we assume that the pay-off function $p_t$ is non-increasing in the variables $\wt$ and $\zt$. The constraint $x_ty_t = 0$ is called the \textit{complementarity} constraint and enforces that the merchant cannot simultaneously buy and sell in the same time period. This constraint is particularly relevant in the context of energy markets. The constraint $y_t\leq s_{t-1}$ represents the common assumption in the warehouse problem that at any time $t$, all sales precede purchases. In the single vendor case, these constraints are implied by  the complementarity constraints.

In this paper we extend this model in two ways. First, we allow the merchant to interact with multiple vendors in the market. Let $\vset = \{1,2,\ldots,\vbound\}$ be the set of vendors. Let $\Lxvt$ and $\Uxvt$ be the lower and upper bounds on purchase quantities when purchasing from vendor $v$ during time period $t$ and let $\Lyvt$ and $\Uyvt$ be the lower and upper bounds on sales quantities. We refer to the lower (upper) bounds as minimum (maximum) capacities. Let continuous variables $\xvt$ and $\yvt$ denote the amount purchased (sold) from (to) vendor $v$ during time period $t\in \tset$. In addition, for each $t\in \tset$ and $v\in\vset$ we define a binary variable $\wvt$ which can take value zero only if no purchases were made from vendor $v$ during time period $t$, and a binary variable $\zvt$  which can take value zero only if no sales were made to vendor $v$ during time period $t$. We will use the shorthand $x_{\vset,t}$ to denote the set of variables $\{x_{v,t}\}_{\vset}$. The pay-off function now takes as input all of these vendor dependent variables at time $t$ and so the objective becomes $\max \mathbb{E}_R\left[\sum_{t\in\tset} \pt(\x_{\vset,t},\y_{\vset,t},\sst,\w_{\vset,t},\z_{\vset,t} \;|\; \Rt)\right]$.

Our second extension is to allow more general complementarity constraints. 
For any pair of time periods $t_2,t_1\in\tset$ such that $t_2\geq t_1$, we consider a collection of
complementarity constraints $\compset{t_1}{t_2}$ 
that contain 4-tuples $(V_1,V_2,B_1,B_2)$ where $V_1$ is one of the variables $\x_{v,t_1}$ or $\y_{v,t_1}$ for some $v\in\vset$ and $V_2$ is one of the variables $\x_{v,t_2}$ or $\y_{v,t_2}$ for some (possibly different) $v\in\vset$. $\cboundone$ is either  $0$, or it is equal to one of the lower or upper bounds of the variable $\cvarone$. $\cboundtwo$ is similar.
Such a 4-tuple corresponds to the constraint $(V_1-B_1)(V_2-B_2) = 0$. For example, if the set $\compset{t_1}{t_2}$ contains a 4-tuple $(x_{1,t_1},x_{3,t_2}, U^{x}_{1,t_1}, 0)$, then the corresponding constraint would be $(x_{1,t_1} = U^{x}_{1,t_1})$ OR  $(x_{3,t_2} = 0)$. With this notation, the extended model can be formally described as:
\begin{align*}\label{problem}
    \max & \;\;\;\;\mathbb{E}_R\left[\sum_{t\in\tset} \pt( \x_{\vset,t},\y_{\vset,t},\sst,\w_{\vset,t},\z_{\vset,t}  \;|\; \Rt)\right]\tag{WP}\\
    \text{s.t.}\quad& (\x,\y,\sst,\w,\z)  \in Q
\end{align*}
where the feasible region $Q$ is defined as
\begin{align}\label{set:Q}
    &\;\;\;\; \sst = \s_{t-1} - \sum_{v\in\vset}\yvt + \sum_{v\in\vset}\xvt & \forall\; t\in\tset\tag{$Q$}\\
    &\;\;\;\;\Lst \leq \sst \leq \Ust &\forall\; t\in\tset\notag\\
    &\;\;\;\; \sum_{v\in\vset}\yvt \leq s_{t-1} &\forall\; t\in\tset\notag\\
    &\;\;\;\; (\cvarone-\cboundone)(\cvartwo-\cboundtwo) = 0 & \forall\; t_1\leq t_2\in\tset, (\cvarone,\cvartwo,\cboundone,\cboundtwo)\in \compset{t_1}{t_2}\notag\\
    &\;\;\;\;\Lyvt \zvt \leq \yvt\leq \Uyvt\zvt,& \forall\; t\in\tset, v\in\vset \notag\\
    &\;\;\;\;\Lxvt \wvt\leq \xvt\leq \Uxvt\wvt& \forall\; t\in\tset, v\in\vset \notag\\
    &\;\;\;\; \wvt,\zvt \in \{0,1\} &\forall\; t\in\tset, v\in\vset\notag
\end{align}
Generally the pay-off function $\pt$ is assumed to be linear. In this paper, we only assume that the pay-off function $\pt$ is a convex function for any realization of the random variable $\Rt$. As alluded to before, a common approach to tackle such stochastic problems is to use deterministic solutions with re-optimization at successive time intervals. Hence, we will restrict our attention to solving the deterministic version of Problem \eqref{problem}. Note crucially that the expectation of a convex function remains convex, and so in each time interval, we can replace the stochastic functions $p_t(\;\cdot\;|\;R_t)$ with its expectation and optimize for the resulting deterministic problem with convex objective.

\section{Extreme Points of the Feasible Region} 

In this section, we analyze the extreme points of the convex hull of  feasible solutions $P=conv(Q)$. Using this characterization, we will then construct a network flow problem to find an optimal extreme point solution to the Warehouse Problem  \eqref{problem}. We first analyze the values stock variables can take at an extreme point, and then analyze the possible values for every other variable in each time interval separately. Let 

   \begin{align}\label{def:St}
    S =   \Big\{&K-\sum_{t\in\tset}\sum_{v \in \vset} (\lambda^1_{v,t} L^{y}_{v,t} + \lambda^2_{v,t} U^{y}_{v,t} + \lambda^3_{v,t} L^{x}_{v,t} + \lambda^4_{v,t}U^{x}_{v,t})\::\:
    \nonumber\\&\quad K\in \{0,\snought,\Us_1,\ldots,U^s_T,L^s_1,\ldots,L^s_T\},
     \\
     &\quad \lambda^j_{v,t} \in \{-1,0,1\}, \;\forall j\in\{1,2,3,4\}, ~t\in\tset, ~v\in\vset\Big\}\nonumber
    \end{align}

The proof of the following two lemmas can be found in Appendix \ref{appendix:lemma1proof}.
\begin{lemma}\label{thm:extstock}
For any extreme point of the polytope $P$, we have $s_t\in S$ for all time periods $t\in\tset$.
\end{lemma}

Lemma \ref{thm:extstock} establishes that the values of the stock variables in extreme points of $P$ come from a set of size at most $T3^{VT}$. We now provide a description of the values that the production and sales variables $x$ and $y$ can take in an extreme point of the polytope $P$. We next argue  that for an extreme point $(\x^*,\y^*,\w^*,\z^*,\s^*)$ of $P$, the possible values for $x^*_{\vset,t}$ and $y^*_{\vset,t}$ variables can be characterized using the values of $s^*_{t-1}$ and $s^*_t$ only. 


\begin{lemma}\label{thm:extxy}
     Let $s'$ and $s''$ be two stock levels in the set $S$, and let $q = (\x^*_{\vset,t},\y^*_{\vset,t},\w^*_{\vset,t},\z^*_{\vset,t},\sst^*)$ be an extreme point of the polytope $P$ with $s_t^* = s'$ and $s_{t-1}^* = s''$. Then the value of the vector $(x_{\vset,t}^*,y_{\vset,t}^*)$ lies in a set of size $O(V^23^{2V})$.
 \end{lemma}
As we discuss in Section \ref{sec:log} in more detail, the size of this set is obtained by finding an upper bound on the number of possible extreme point solutions to a warehouse problem with one time period and fixed initial and final inventory levels. The size of this set is relevant in the algorithm we present in the next section as we solve an optimization problem over this set via enumeration. However, assumptions like linear per-unit costs \cite{devalkar} can help solve this optimization problem more efficiently.

Having provided a characterization of the extreme points of the polytope $P$, we now construct a network such that solving a shortest path problem in this network gives an optimal solution to the warehouse problem \ref{problem}. Note that by lemma \ref{thm:extstock} and lemma \ref{thm:extxy}, the total number of extreme points of the polytope $P$ could be as large as $|S|^T(V^23^{2V})^{T}$  where $S$ was defined in equation \eqref{def:St}. Note that this is not only exponential in the number of vendors but also exponential in the number of planning periods.

\section{Network Flow Formulation}

The main idea of the network flow formulation is borrowed from the works of Wolsey and Yaman \cite{wolsey} and Bansal and Gunluk \cite{bansal2023}. Having found the values of stock levels in extreme points and the corresponding values of purchase and sales quantities, one can construct a network whose nodes correspond to these stock levels and edges correspond to the values of the purchase and sales variables. Every path in this network corresponds to a feasible point of the warehouse problem and every extreme point of the convex hull of the feasible region of the warehouse problem corresponds to a path in the constructed network. In our model, the construction of such a network is more nuanced due the generalized complementarity constraints as not all paths correspond to feasible solutions. We will therefore create multiple nodes for each stock level and time period and augment each such node with information on complementarity constraints that are yet to be satisfied. To formalize this idea, at any time $t\in \tset$, define the set of relevant constraints to be the complementarity constraints that have conditions on both a past and a future time-period, including time $t+1$.
\[\mathcal{C}^{r}_t = \bigcup_{t_1\leq t+1\leq t_2}\mathcal{C}_{t_1,t_2}
\]
and define the \textit{thickness} of the set of generalized complementarity constraints to be
\[
\hat{C} = \max_{t\in\tset}\big\{\,|\mathcal{C}^{r}_t|\,\big\}
\]
We also need a definition of relevant constraints that would not remain relevant in the next time interval i.e. complementarity constraints where $t_2 = t+1$. Define the set of deadline constraints to be
\[
\mathcal{C}^{d}_t = \bigcup_{t_1\leq  t+1}\mathcal{C}_{t_1,t+1}
\]

We can now construct a directed acyclic network $G=(N,A)$ as follows: The node set $N = \bigcup_{t=0}^TN_t$ where $N_t$ has a node for every $(s,C)$ where $s\in S$ and $C\subseteq \mathcal{C}^{r}_t$. The idea is that any node $(s,C) \in N_t$ keeps track of the stock level $s$ at time $t$ and the set of relevant constraints $C$ that have not yet been satisfied. We then construct $A$ by adding (parallel) arcs from $(s,C)$ in $N_{t-1}$ to $(s',C')$ in $N_t$ corresponding to the values $(x_{\vset,t},y_{\vset,t},w_{\vset,t},z_{\vset,t})$ can take in extreme points based on the values of $s$ and $s'$ (these were enumerated in Theorem \ref{thm:extxy}) such that

\begin{enumerate}
    \item[(i)] The values of $x_{\vset,t}$ and $y_{\vset,t}$ satisfy every deadline constraint in $C\cap \mathcal{C}^{d}_{t-1}$.\\[.05cm]
    \item [(ii)] $C' = (C \backslash C_{sat})\cup C_{unsat}$ where $C_{sat}$ are all constraints in $C$ satisfied by $x_{\vset,t}$ and $y_{\vset,t}$ and $C_{unsat}$ are all constraints in $\mathcal{C}^{r}_{t} - \mathcal{C}^{r}_{t-1}$ that are unsatisfied by $x_{\vset,t}$ and $y_{\vset,t}$.
\end{enumerate}

By maintaining condition (i), we will ensure that every generalized complementarity constraint is necessarily satisfied. Condition (ii) ensures that we are keeping track of the correct set of relevant constraints that are yet to be satisfied. 

Finally, for an arc $a$ corresponding to particular values of $(x_{\vset,t},y_{\vset,t},w_{\vset,t},z_{\vset,t})$, connecting a node corresponding to $s_{t-1}$ to a node corresponding to $s_t$, set its weight to be the immediate pay-off $p_t(x_{\vset,t},y_{\vset,t},w_{\vset,t},z_{\vset,t},s_t)$. With this construction we have the following theorem:

\begin{theorem}\label{thm:runningtime}
    The warehouse problem \eqref{problem} can be solved in $O(T V^2 3^{2V}|S|^22^{2\hat{C}})$ time by solving a longest path problem in the (acyclic) graph $G$.
\end{theorem}

\begin{proof}
    First observe that the network $G$ constructed above has the following properties:
    \begin{enumerate}
    \item Every path from $N_0$ to $N_T$ in $G$ corresponds to a a feasible point $(x,y,w,z,s)$ in $Q$ and the length of the path is equal to $\sum_{t\in\tset}p_t(x_{\vset,t},y_{\vset,t},w_{\vset,t},z_{\vset,t},s_t)$.\\[.05cm]
    \item For every extreme point of $P$, there exists a path from $N_0$ to $N_T$ in $G$ corresponding to the extreme point.
\end{enumerate}
The first property is true since a path from $N_0$ to $N_T$ moves from a vertex at time $t$ corresponding to an inventory level $s_t$ to a vertex at time $t+1$ corresponding to an inventory level $s_{t+1}$ using a feasible set of $x_{\vset,t}$ and $y_{\vset,t}$ variables. The variables $w$ and $z$ can then be set optimally as follows: $w_{v,t} = 0$ if and only if $x_{v,t} = 0$ and $z_{v,t} = 0$ if and only if $y_{v,t} = 0$. Recall that the pay-off function is non-increasing in the variables $w$ and $z$ and so the above assignment of these variables is indeed optimal. The second property holds true by the construction of the network $G$.

Due to the above two properties and the assumption that the objective function is convex, a longest path from $N_0$ to $N_T$ in $G$ corresponds to an optimal solution to the warehouse problem \eqref{problem}. The size of the network flow formulation $G=(N,A)$ is bounded as follows: the number of nodes $|N| \leq T|S|^22^{2\hat{C}}$ and the number of arcs $|A| \leq T V^2 3^{2V}|S|^22^{2\hat{C}}$ by lemma \ref{thm:extxy}. Since a longest path in a directed acyclic graph can be calculated in time $O(|N|+|A|)$, the running time of finding a longest path in $G$ is bounded by $O(T V^2 3^{2V}|S|^22^{2\hat{C}})$. \qed
\end{proof}

We will refer to the above algorithm of finding a longest path in the network $G$ and outputting the corresponding feasible solution in the set $Q$ as the \textit{longest path algorithm}. Even though the running time of the longest path algorithm is  exponential, we will consider realistic conditions in the next section that make it poly-time.



\section{Polynomially Solvable Cases and NP-Hardness}

In this section, we list conditions which together ensure that the longest path algorithm has polynomial running time. We also note that the absence of some of these conditions will render the problem NP-Hard and provide the reductions therein. Recall from Theorem \ref{thm:runningtime} that the running time of the longest path algorithm is $O(T V^2 3^{2V}|S|^22^{2\hat{C}})$. For this to be polynomial, we will have to control the sizes of $|S|,V$ and $\hat{C}$.

\subsection{Bounds From a Lattice} 

As observed in Section 6 of \cite{bansal2023}, the warehouse problem is NP-Hard even in the single vendor case when the bounds on purchase and sales quantities are allowed to be arbitrary. Hence similar to \cite{bansal2023}, we need the condition that all bounds on purchase and sales quantities lie on a suitable lattice.

\begin{assume} \label{A1} For all $v\in\vset$ and $t\in\tset$ assume that
\begin{align*}
    L^y_{v,t}, U^y_{v,t}, L^x_{v,t}, U^{x}_{v,t} \in \left\{\sum_{i=1}^k \alpha_id_i : \alpha_i \in \mathbb{Z}, |\alpha_i|\leq \gamma\right\}
\end{align*}
where $k, d_1,.,d_k$ are fixed positive constants and $\gamma = O(T^c)$ for some constant $c$.
\end{assume}

Under this assumption, all lower and upper bounds of the  $\xvt$ and $\yvt$ variables belong to a set of size $O(T^{kc})$. Moreover, the set $S$ containing all possible stock levels in extreme points, described in Equation \eqref{def:St} now become  a subset of $\bar S$, where
\begin{align*}
    \bar S~=~\left\{K+\sum_{i=1}^k \beta_i d_i : \beta_i \in [-VT\gamma,VT\gamma]\cap\mathbb{Z}, K\in\{0,s_0,U^s_1,\ldots,U^s_T,L^s_1,\ldots,L^s_T\}\right\}.
\end{align*}
Note that  $|\bar S|\le (2T+2)(2VT\gamma+1)^k$.
Hence, the size of the set $S$ can be bounded by $|\bar S| = O(T(2VT\gamma)^k) = O(T^{kc+k+1}V^{k})$. 

The assumption that the bounds come from a lattice are not very restrictive and are of practical importance. The bounds on purchase/production and sales quantities come about from operational considerations and it is unrealistic to consider bounds that drastically and randomly change again and again. Some examples captured in this setting are (i) If there are a constant number of levels of purchase/production and sales, say \textit{low, medium} or \textit{high}. (ii) If the bounds are all linear combinations of some basic units of purchase/production and sales quantities. For instance, by choosing the values of $d_i$ to be $10^i$, one can capture all bounds up to say a million using a constant sized basis.

If the bounds on purchases and sales variables can be arbitrary, then the warehouse problem \eqref{problem} is NP-Hard even in the single vendor case \cite{bansal2023}. However, the single-vendor problem becomes poly-time solvable if the bounds are time-independent. In contrast, the warehouse problem with multiple vendors \eqref{problem} is NP-Hard even if each vendor has time-independent bounds.

\begin{lemma}
    The warehouse problem \eqref{problem} where each vendor has (possibly different) time-independent bounds on purchase and sales quantities is NP-Hard
\end{lemma}

\begin{proof}
    We reduce the problem to the single-vendor warehouse problem with time-dependent bounds, shown to be NP-Hard in \cite{bansal2023}. Let $\{L^x_t,U^x_t,L^y_t,U^y_t\}_{t\in\tset}$ be the bounds in an instance of the single-vendor problem. We construct a multi-vendor instance with $V=T$. We identify vendor $i$ with time period $i$ and set its time-independent bounds to be $L^x_i,U^x_i,L^y_i,U^y_i$. The fixed costs for any vendor during time-period $i$ can be set large enough so that it is never optimal to trade with any other vendor during time period $i$, and the costs associated with vendor $i$ can be set to be the same as the single-vendor instance. This completes the reduction. \qed
\end{proof}

\subsection{Logarithmic Number of Vendors} \label{sec:log}

We next state a condition that bounds the number of possible values that  $x_{\vset,t}$ and $y_{\vset,t}$ variables can take in extreme points of $P$ for given values for $s_{t-1}$ and $s_t$.  Recall from Lemma $\ref{thm:extxy}$ that there are $O(V^23^{2V})$ such values. Thus, if we require that $V=O(\log T)$, we can bound $3^{2V}$ by a polynomial in $T$.

\begin{assume}\label{A2} Assume that $V \leq k'\log T$ for some constant $k'$.\end{assume}

Note that the $3^{2V}$ term in the running time of the longest path algorithm is to find the optimal choice of $(x_{\vset,t},y_{\vset,t})$ after having fixed $s_{t-1}$ and $s_t$. The longest path algorithm solves this problem by enumerating over all possible extreme values of $(x_{\vset,t},y_{\vset,t})$. This may not be necessary. Formally, we are trying to solve the following optimization problem where $s_t$ and $s_{t-1}$ are fixed inputs and $p$ is a convex function:

\begin{align*}
    \max\;\;\;\; & p(x_{\vset},y_{\vset},z_{\vset},w_{\vset})\\
    \text{s.t.}\;\;\;\;&\;\;\;\; \sst - \s_{t-1} = \sum_{v\in\vset}\x_v - \sum_{v\in\vset}\y_v\\
    &\;\;\;\; \sum_{v\in\vset}\y_v \leq s_{t-1} \\
    &\;\;\;\;L^y_v \z_v \leq \y_v\leq \Uy_v\z_v,& v\in\vset \notag\\
    &\;\;\;\;L^x_v \w_v\leq \x_v\leq \Ux_v\w_v& v\in\vset \notag\\
    &\;\;\;\; \w_v,\z_v \in \{0,1\} & v\in\vset\notag
\end{align*}

The above problem can be viewed as the warehouse problem \eqref{problem} with just a single time-step. If this problem itself is NP-Hard, then the more general warehouse problem \eqref{problem} will also be NP-Hard. 
Note that the above problem can model any box-constrained convex maximization problem since we can set the constant $s_{t-1}$ to be arbitrarily large, making the second constraint redundant. This problem is known to be NP-Hard. Hence some conditions have to be imposed for the problem to be poly-time solvable. The condition we impose (i.e. logarithmic number of vendors) is not very restrictive, since in most practical settings, the number of vendors that a merchant considers trading with in any time period is a constant sized set.

\subsection{Logarithmic Thickness} 

Lastly, we state a condition that bounds the number of relevant constraints that the longest path algorithm keeps track of. Recall that the number of subsets of relevant constraints is bounded by $2^{2\hat{C}}$ where $\hat{C}$ is the thickness of the set of generalized complementarity constraints. Thus, if we require that $\hat{C}  = O(\log T)$, we can bound $2^{2\hat{C}}$ by a polynomial in $T$.

\begin{assume}\label{A3} Assume that $\hat{C} \leq k'\log T$ for some constant $k'$.\end{assume}

This condition again is not very restrictive. It is unlikely that the decision to purchase/sell at a particular time period, affects the decision to purchase/sell at a much later time period. It is nonetheless unclear if such a condition is required and whether the removal of this condition will make the problem NP-Hard.

Combining the above three assumptions is sufficient to bound the running time of the longest pat algorithm and we obtain the following theorem.

\begin{theorem}
    Under Assumptions \ref{A1}, \ref{A2}, and \ref{A3}, the warehouse problem \eqref{problem} can be solved in polynomial time using the longest path algorithm.
\end{theorem}

\section{Modeling Practical Constraints}

In this section, we justify our motivation for considering multiple vendors and generalized complementarity constraints. We show that practical constraints relevant in energy markets, and others, can be captured in this more general framework.

\subsection{Discounted Costs and Piece-wise Convex Pay-offs}

The results in \cite{bansal2023} deal with the single vendor warehouse problem when the pay-off is convex. However, they are unable to model practical considerations such as discounted or surged costs. For instance, the per-unit cost for purchases could be $c_1$ up to a certain quantity of sales, after which a surged rate kicks in and the per-unit cost becomes $c_2 > c_1$. This is precisely how consumers on the grid are currently charged for electricity consumption \cite{barreto2015incentives}. These lead to piece-wise linear concave objectives and hence cannot be solved by the methods in \cite{bansal2023}. Nonetheless, such constraints can be modelled using multiple vendors and generalized complementarity constraints. For instance, suppose the per-unit cost of purchases at time $t$ is given by,
\begin{align*}
    \begin{cases} c_1, &\text{if  } 0\leq x_t \leq U_1 \\
    c_2, &\text{if  } U_1<x_t\leq U_2 \\
    c_3, &\text{if}U_2<x_t\leq U_3\end{cases}
\end{align*}
We can then introduce three vendors $\{1,2,3\}$ with the following constraints:
\begin{align*}
    & x_{1,t} \leq U_1,\quad x_{2,t} \leq U_2-U_1,\quad x_{3,t} \leq U_3-U_2\\[.05cm]
    & (x_{1,t} - U_1)x_{2,t} = 0\\[.05cm]
    & (x_{2,t}-(U_2-U_1))x_{3,t} = 0
\end{align*}
and we can set the per-unit cost of purchases from vendor $i$ as $c_i$. The generalized complementarity constraints above ensure that $x_{2,t} > 0$ implies $x_{1,t} = U_1$ and $x_{3,t}> 0$ implies $x_{2,t} = U_2-U_1$. The above method can also be used to model any piece-wise continuous convex pay-off function.

\subsection{Ramp-up/Ramp-down Constraints}

In practice, the commodity being traded is not only purchased, but could also be produced. For example the merchant could have a generator that produces electricity or a factory/mill that manufactures the commodity. The generator could have various power settings that determine the speed at which electricity is produced. However, it is often not possible to ramp-up/ramp-down these power settings drastically in consecutive time periods \cite{wang2016value}. Such constraints can be modelled using multiple vendors and generalized complementarity constraints. For instance, suppose the generator has three power settings $\textit{low,medium}$ and $\textit{high}$ and suppose it is not possible to switch from $\textit{low}$ to $\textit{high}$ or vice-versa during consecutive time periods, then we can introduce three vendors $\{l,m,h\}$ corresponding to the three power settings and add generalized complementarity constraints of the form,

\begin{align*}
    &x_{l,t} x_{m,t} = 0,\quad x_{l,t}x_{h,t} = 0,\quad x_{m,t}x_{h,t} = 0\\[.05cm]
    &x_{l,t}x_{h,t+1} = 0,\quad x_{h,t}x_{l,t+1}=0
\end{align*}

The first constraint ensures that only one power setting is being used at any time. The second constraints ensures the ramp-up/ramp-down properties.

\subsection{Production and Sales in Batches}\label{subsec:batchpricing}

In some markets, the cost of production and sales occurs in batches. For instance, the major cost could be associated with transporting the goods and not producing them. Thus the cost is directly proportional to the number of trucks being used and not the amount of goods being produced. This has been modelled as production and sales in batches in lot-sizing problems \cite{li2004dynamic}. Such batch pricing could also occur due to contractual agreements. To model batch-pricing, we introduce multiple vendors to represent the number of batches being produced. We do so in a logarithmic scale to ensure a polynomial number of new variables. Let $U$ be the batch size and $2^k-1$ be the maximum number of batches that can be purchased. Let $c_t$ be the cost per batch at time $t$. We introduce $k$ vendors at each time period $t$, $x_{0,t},\ldots,x_{k-1,t}$ where $x_{i,t}$ corresponds to the purchase of $2^i$ batches. We set the fixed cost of vendor $i$ at time $t$ to be $2^ic_{t}$ and the per-unit cost to be zero. The bounds for vendor $i$ at time $t$ are simply given by $2^i U$. If the maximum number of batches that can be purchased is not of the form $2^k-1$, then we can add more complementarity constraints to ensure that the purchased amount is correctly bounded, as outlined in the Appendix \ref{subsec:multiplevariables}.

\section{Further Extensions}

Our methods and results can further be extended to capture useful generalizations of the complementarity constraints which in the current model \eqref{problem} are conditions of the form $(V_1 = B_1)$ OR $(V_2 = B_2)$. We can consider an OR of multiple conditions, some of which could be inequalities instead of equalities. In addition, the bounds $B_1$ and $B_2$ need not be restricted to zero, lower and upper bounds of variables. The descriptions of these extensions are provided in the Appendix \ref{sec:furterextensions}.

\bibliographystyle{splncs04.bst}
\bibliography{IPCO2024}
\newpage
\appendix
\section{Description of Further Extensions}\label{sec:furterextensions}

In this section, we point out some further extensions that are easily obtainable by following the methods laid out in this paper. In the current warehouse model \eqref{problem}, the complementarity constraints are essentially conditions of the form $(V_1 = B_1)$ OR $(V_2 = B_2)$. We can extend these conditions in various ways:

\subsection{Multiple Variables}\label{subsec:multiplevariables}

We can consider OR conditions between multiple constraints $\{V_i = B_i\}_{i=1}^k$. Such constraints can model practical considerations like down-times. Generators in electricity markets heat up and need to be shut down at regular intervals. Thus, we require constraints of the form $x_{v,1} = 0$ OR $x_{v,2} = 0$ OR $x_{v,3} = 0$ OR $x_{v,4} = 0$ (If the down-time is once every 4 time intervals for example). Hence, complementarity constraints of the form $\prod_{i=1}^k (V_i - B_i) = 0$ can be useful. All of our results hold under this extension as well. The thickness of the set of complementarity constraints $\hat{C}$ is still defined as the maximum number of relevant constraints at any time-period.

Such complementarity constraints help model batch-pricing mentioned in Section \ref{subsec:batchpricing}. We add complementarity constraints between variables that would exceed the maximum number of batches that can be purchased. For instance, if the maximum number of purchasable batches is 10, then our variables are $x_0,x_1,x_2,x_3$ corresponding to purchases of $1,2,4,8$ batches respectively. We can add the complementarity constraints $x_3x_2 = 0$ and $x_3x_1x_0 = 0$ to ensure at most 10 batches are purchased.

\subsection{Inequalities}

It might be useful to consider complementarity constraints of the form $(V_1 \geq B_1)$ OR $(V_2 \leq B_2)$ (and other combinations of the constraints $\leq,=,\geq$). These can be modelled by adding additional variables. We add the constraint $(V_1 - B_1 - \ell_1)(V_2 - B_2 + \ell_2) = 0$ where $\ell_1,\ell_2\geq 0$ are new slack variables we introduce into the model with the only constraint that they are non-negative. With a few modifications to our proofs, we can show that the set $S$ defined in Equation \eqref{def:St} still contains all possible extreme values of the stock variables. In addition, Lemma \ref{thm:extstock} and Lemma \ref{thm:extxy} still hold. Hence, such problems can also be solved by finding the longest path in a suitably constructed network.

\subsection{Different Bounds}
In the current model, the constants $B_i$ that appear in complementarity conditions are required to be either zero or the lower bound of variable $V_i$ or the upper bound of variable $V_i$. We can relax this condition on $B_i$ and allow these constants to be any non-negative number. However, the description of the inventory levels in extreme points given by set $S$ in Equation \eqref{def:St} would now have terms corresponding to the new values of the constants $B_i$.
\section{Omitted proofs}
\subsection*{Proof of Lemma \ref{thm:extstock}}\label{appendix:lemma1proof}

\begin{proof}
    The proof is by contradiction. Let $q^* = (\x^*,\y^*,\w^*,\z^*,\s^*)$ 
    be an extreme point  of $P$ and assume that $s^*_{\tau} \not\in S$ for some $\tau\in\tset$. 

    If for all $t\le\tau$ and for all $v\in\vset$ we have $\x^*_{v,t} \in\{0,L^x_{v,t},U^x_{v,t}\}$ and $\y^*_{v,t} \in\{0,L^y_{v,t},U^y_{v,t}\}$, then $s^*_{\tau}$ is obtained by starting from $s_0\in K$ and always purchasing and selling at either maximum or minimum capacity, or at zero, for all previous time periods and  $s^*_{\tau}\in S$. 
    Therefore, there exists some $t\le\tau$ and $v\in\vset$ such that either $\x^*_{v,t} \not\in\{0,L^x_{v,t},U^x_{v,t}\}$ or $\y^*_{v,t} \not\in\{0,L^y_{v,t},U^y_{v,t}\}$. 
    Let $t_1$ be the largest such $t\leq \tau$ and $v_1\in\vset$ be the associated vendor.
     We will next assume that $L^x_{v_1,t_1}<\x^*_{v_1,t_1}<U^x_{v_1,t_1}$ and argue that $q^*$ cannot be an extreme point. The case when $L^y_{v_1,t_1}<\y^*_{v_1,t_1}<U^y_{v_1,t_1}$ is handled similarly. 
     
     First note that for all time periods $t$ such that $t_1\leq t \leq \tau$,  the stock levels must be strictly between its bounds i.e.
    \begin{align}\label{property1}
        L^s_t < s^*_t < U^s_t && \forall t_1\leq t \leq \tau
    \end{align} 
    Otherwise, $s^*_{\tau}$ can be obtained by starting from $s^*_t \in \{L^s_t, U^s_t\}$ and always purchasing and selling at either maximum or minimum capacity, or at zero, implying that $s^*_{\tau} \in S$. 
    
    Also note that, for any time period $t$ such that $t_1\leq t < \tau$, the total amount sold to all vendors cannot equal the current stock i.e.
    \begin{align}\label{property2}
        s^*_t > \sum_{v\in \vset}y^*_{v,t+1}  && \forall t_1\leq t < \tau.
    \end{align}
    Otherwise, $s^*_{\tau}$ is obtained by starting from $s^*_t - \sum_{v\in \vset}y^*_{v,t+1} = 0$ and  always purchasing and selling at capacity, or at zero, implying $s^*_{\tau} \in S$. 
    
    Next, let $t_2$ be the smallest time period $t\geq \tau$ such that $s^*_{t} \in \{L^s_{t}, U^s_{t}\}$ or $s^*_{t} = \sum_{v\in\vset}y^*_{v,t+1}$. If no such $t$ exists, then due to equations \eqref{property1} and \eqref{property2}, we can increase and decrease $x^*_{v_1,t_1}$ (and all stock levels $\{s_t\}_{t\geq t_1})$ by a small quantity $\epsilon>0$ to obtain two distinct points in $P$. This contradicts the assumption that $q^*$ is an extreme point, as it can be now be expressed as a convex combination of these points. 
    We next consider the following two cases.

    \textbf{Case 1:} $s^*_{t_2} \in \{L^s_{t_2},U^s_{t_2}\}$. Suppose, there exists a time period $\tau<t\leq t_2$ such that for some vendor $v'$ either $x^*_{v',t} \not\in \{0,U^x_{v',t},L^x_{v',t}\}$ or $y^*_{v',t} \not\in \{0,U^y_{v',t}, L^y_{v',t}\}$, then due to equations \eqref{property1} and \eqref{property2}, we can increase and decrease $x^*_{v_1,t_1}$ and $x^*_{v',t}$ (or $y^*_{v',t}$) by a small quantity $\epsilon>0$ (and update the stock variables accordingly) to obtain two distinct points in $P$. This would contradict the assumption that $q^*$ is an extreme point. Hence there is no such time period $t$. But now, $s^*_{\tau}$ is obtained by starting from $s^*_{t_2} \in \{L^s_{t_2}, U^s_{t_2}\}$ and moving back in time, only purchasing and selling at zero/min/max capacity. Thus $s^*_{\tau} \in S$ (as the coefficients $\lambda$ in the definition of the set $S$ can take values in $\{-1,0,1\}$).

    \textbf{Case 2:} $s^*_{t_2} = \sum_{v\in\vset}y^*_{v,t_2+1}$. Similar to the previous case, we observe that if there exists a time period $\tau<t\leq t_2$ such that for some vendor $v'$ either $x^*_{v',t} \not\in \{0,U^x_{v',t},L^x_{v',t}\}$ or $y^*_{v',t} \not\in \{0,U^y_{v',t}, L^y_{v',t}\}$, or if $y^*_{v',t_2+1} \not\in\{0,U^y_{v',t_2+1},L^y_{v',t_2+1}\}$, then we can show that $q^*$ is not an extreme point. But now, $s^*_{\tau}$ is obtained by starting from $s^*_{t_2} - \sum_{v\in\vset}y^*_{v,t_2+1} = 0$ and moving back in time, only purchasing and selling at zero/min/max capacity. Thus $s^*_{\tau} \in S$. 
    
    This completes the proof when  $L^x_{v_1,t_1}<\x^*_{v_1,t_1}<U^x_{v_1,t_1}$. The case when $L^y_{v_1,t_1}<\y^*_{v_1,t_1}<U^y_{v_1,t_1}$ is handled the same way. Note that the perturbations we make to feasible solutions do not violate the complementarity constraints as we only perturb variables that are nonzero, and strictly between their lower or upper bounds.\qed 
\end{proof}

\subsection*{Proof of Lemma \ref{thm:extxy}}\label{appendix:lemma2proof}

\begin{proof}
We will consider two cases and argue that almost all purchase and sales variables $(x^*_{v,t},y^*_{v,t})_{v\in\vset}$ must be either at zero, or at minimum or maximum capacity. 

    \textbf{Case 1:} $s_{t-1}^* > \sum_{v\in\vset}y_{v,t}^*$. In this case, we observe that at most one of the variables in $(x^*_{v,t},y^*_{v,t})_{v\in\vset}$ can be strictly between its minimum and maximum capacity as otherwise we can increase and decrease these variables suitably by a small quantity $\epsilon>0$ to express $q^*$ as a convex combination of distinct  points in $P$. There are $2V$ choices for the variable with value strictly between its minimum and maximum capacity, say $x^*_{v,t}$ (or, $y^*_{v,t}$). Every other variable has three possible values: zero, minimum capacity or maximum capacity. Having fixed these variables, the value of $x^*_{v,t}$ (or, $y^*_{v,t}$) is determined by the equation $s_t^* = s_{t-1}^* - \sum_{v\in\vset}y^*_{v,t} + \sum_{v\in\vset}x^*_{v,t}$. Thus the total number of possible values for $(x^*_{v,t},y^*_{v,t})_{v\in\vset}$ is at most $2V\cdot 3^{2V-1}$.

    \textbf{Case 2:} $s_{t-1}^* = \sum_{v\in\vset}y^*_{v,t}$. In this case at most one of the sales variables $y^*_{v,t}$  can be strictly between its minimum and maximum capacity. This holds because otherwise, we can again express $q^*$ as a convex combination of distinct feasible points obtained by increasing and decreasing these sales variables  by a small number $\epsilon>0$. There are $V$ choices for the sales variable that is strictly between its minimum and maximum capacity, say $y^*_{v,t}$. Remaining sales variables have three possible values: zero, minimum capacity or maximum capacity. Having fixed these variables, the value of $y^*_{v,t}$ is determined by the equation $s_{t-1}^* = \sum_{v\in\vset}y^*_{v,t}$. The purchase variables $x^*_{v,t}$ have a similar analysis. Thus the total number of possible values for $(x^*_{v,t},y^*_{v,t})_{v\in\vset}$ is at most $V^23^{2V-2}$. 
    
    This completes the proof. Note that the perturbations we make to feasible solutions do not violate the complementarity constraints. \qed
\end{proof}

\end{document}